\documentclass [prl, superscriptaddress, showpacs, twocolumn ]{revtex4}
\usepackage{graphicx}
\usepackage{dcolumn}

\newcommand{\grad}{$^{\circ}$}
\newcommand{\CaFeAs}{CaFe$_4$As$_3$}
\newcommand{\FeAs}{Fe$_2$As$_2$}

\begin{document}

\title{Incommensurate spin-density wave and magnetic lock-in transition in \CaFeAs\ }
\author{P. Manuel}
\affiliation{ISIS facility, STFC Rutherford Appleton
Laboratory, Chilton, Didcot, Oxfordshire, OX11 0QX, United Kingdom.}
\author{L. C. Chapon}
\affiliation{ISIS facility, STFC Rutherford Appleton
Laboratory, Chilton, Didcot, Oxfordshire, OX11 0QX, United Kingdom.}
\author{I. S. Todorov}
\affiliation{Materials Science Division, Argonne National Laboratory, Argonne, IL 60439, USA}
\author{D. Y. Chung}
\affiliation{Materials Science Division, Argonne National Laboratory, Argonne, IL 60439, USA}
\author{J.-P. Castellan}
\affiliation{Materials Science Division, Argonne National Laboratory, Argonne, IL 60439, USA}
\author{S. Rosenkranz}
\affiliation{Materials Science Division, Argonne National Laboratory, Argonne, IL 60439, USA}
\author{R. Osborn}
\affiliation{Materials Science Division, Argonne National Laboratory, Argonne, IL 60439, USA}
\author{P. Toledano}
\affiliation{Laboratory of Physics of Complex Systems, University of Picardie, 33 rue Saint-Leu, 80000 Amiens, France}
\author{M. G. Kanatzidis}
\affiliation{Materials Science Division, Argonne National Laboratory, Argonne, IL 60439, USA}
\affiliation{Department of Chemistry, Northwestern University, Evanston, IL 60208, USA}
\date{\today}

\begin{abstract}
The magnetic structure for the newly discovered iron-arsenide compound \CaFeAs\ has been studied by neutron powder diffraction. Long-range magnetic order is detected below 85K, with an incommensurate modulation described by the propagation vector k=(0,$\delta$,0), $\delta\sim$ 0.39. Below $\sim$ 25K, our measurements detect a first-order phase transition where $\delta$ locks into the commensurate value $\frac{3}{8}$. A model of the magnetic structure is proposed for both temperature regimes, based on Rietveld refinements of the powder data and symmetry considerations. The structures correspond to longitudinal spin-density-waves with magnetic moments directed along the \textit{b}-axis. A Landau analysis captures the change in thermodynamic quantities observed at the two magnetic transitions, in particular the drop in resistivity at the lock-in transition. 
\end{abstract}


\maketitle

\begin{figure}[h!] 
\includegraphics[width=210pt]{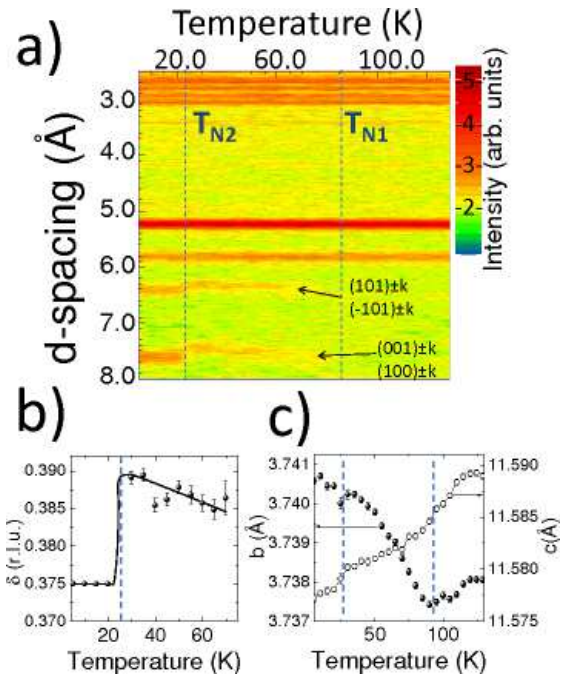} 
\caption{(Color online) a) Neutron thermodiffractograms of \CaFeAs\ collected on the low angle bank (scattering angle 2$\theta$ = 17.98\grad) on the GEM diffractometer. The intensity is color-coded. b) temperature evolution of the component $\delta$ of the magnetic propagation vector k =(0,$\delta$,0). The solid line is a guide to the eye c) temperature evolution of the lattice parameters \textit{b} (closed circles) and \textit{c} (open circles). Dashed lines indicate the transitions T$_{N1}$ (panels a and c) and T$_{N2}$ (all panels).}
\label{fig:tdep}
\end{figure}

\indent The recent discovery of superconductivity at 26K in iron-based layered compounds La[O$_{1-x}$F$_x$]FeAs \cite{ISI:000253951900032} has triggered a new wave of investigations in the field of superconductivity. There are many similarities with the High Temperature Cuprate superconductors (HTC) with the presence of square planar superconducting layers, in this case derived from tetrahedrally-coordinated \FeAs\ units, separated by charge reservoirs. All the undoped non-superconducting systems, including the so-called 122 compounds A\FeAs\ (where A=Ba,Sr,Ca), undergo a tetragonal to orthorhombic transition accompanied by a commensurate antiferromagnetic (AFM) order with an ordered moment less than 1 $\mu_B$ \cite{ISI:000261891200064}. Chemical doping suppresses the structural transition and superconductivity is observed at critical temperatures as high as 55K and 38K in optimally doped SmFeAs(O$_{1-x}$F$_x$) \cite{ISI:000256252600080} and Ba$_{0.6}$K$_{0.4}$\FeAs\ \cite{ISI:000258975100055} respectively.\\
\indent Since the \FeAs\ layer is the key structural motif in the new iron arsenide superconductors, the ability to explore different topologies that include such building blocks is of paramount importance. The recently discovered \CaFeAs \cite{ISI:000265268100019} compound is an exciting member of this family, where the crystal structure is made of infinite \FeAs\ layers along \textit{b} but finite in a perpendicular direction. These \FeAs\ \emph{ribbons}, interconnected by five-coordinated iron atoms, are arranged in a rectangular cross pattern. Magnetic susceptibility measurements \cite{ISI:000265268100019,ISI:000268617500007} reveal the existence of two transitions at T$_{N1}$ $\approx$90K and T$_{N2}$ $\approx$26K, with antiferromagnetic correlations along the \textit{b}-axis at T$_{N1}$. The heat capacity measurements show a $\lambda$-type anomaly only at T$_{N1}$, and an extremely weak change of slope at T$_{N2}$\cite{ISI:000268617500007}. On the other hand, the resistivity measurements show a weak anomaly at T$_{N1}$, but a more pronounced drop at T$_{N2}$ although the magnitude and direction differ in the two studies\cite{ISI:000265268100019,ISI:000268617500007}. Even though such bulk measurements clearly point to a strong coupling between the magnetic order parameter and resistivity, to date there is no scattering studies of the magnetic state in this system and its dependence on temperature.\\
\indent In this communication, we report on the magnetic structure of \CaFeAs\ probed by neutron diffraction experiments on polycrystalline samples. The transition at T$_{N1}$ corresponds to an incommensurate (ICM) longitudinal spin-density wave (SDW) along the \textit{b}-axis with propagation vector k=(0,$\delta$,0) and 0.375$<\delta <$ 0.39. The transition is second-order with the magnetic order parameter transforming as one of the irreducible representation (IR) of the $Pnma$ space group. At T$_{N2}$, the structure locks into a commensurate (CM) state with $\delta$=$\frac{3}{8}$ and the moments remain predominantly aligned along the $b$-axis. The CM-ICM transition is first order and the CM state is stable down to the lowest temperature measured of 1.5K. A phenomenological description of the magnetic phase transition from the paramagnetic state to the ICM and CM state is able to reproduce all the features found in the temperature dependence of physical quantities. In particular, the sudden drop of electrical resistivity at T$_{N2}$ associated with a surprisingly weak anomaly in the heat-capacity are both explained by the presence of a sixteenth degree invariant that stabilizes the CM state.\\ 
\indent A 0.8g sample of crushed single crystals grown in a Sn flux \cite{ISI:000265268100019} was loaded into a standard 6mm vanadium can and cooled inside a cryostat for neutron experiments performed at the ISIS pulsed neutron source of the Rutherford Appleton Laboratory, U.K.. The temperature dependence of the crystal and magnetic parameters was followed using the General Materials Diffractometer (GEM). Data were collected on cooling between 130K and 5K in 5K steps. Higher-resolution data at low-Q (scattering vector) were collected on the WISH diffractometer on the ISIS Target Station 2 at 1.5K, 30K and 100K. Rietveld refinements were carried-out with the program FullProf \cite{ISI:A1993ME99200007}. Symmetry (representation) analysis is presented using Kovalev's notation \cite{rep:kovalev}.\\
\begin{table}[h!]
\begin{tabular}{c c | c c |c c }
 & & 30K & & 1.5K\\
Site $i$ & Pos. & M$_i$($\mu_B$) & $\Phi_i$ & M$_i$($\mu_B$) & $\Phi_i$ \\
\hline
Fe$_1$ & x=0.0210(5) & 1.4(11) & 0 & 2.14(13) & 0 \\
 & z=0.3135(4) & & & & \\
Fe$_2$ & x=0.0671(4)  & 1.61(14) & 0.14(3) & 1.55(16) & 0.13(2) \\
& z=0.5376(5) & & & & \\
Fe$_3$ & x=0.3037(4)  & 1.67(20) & 0.45(3) & 1.83(8) & 0.56(4) \\
& z=0.1243(4) & & & & \\
Fe$_4$ & x=0.3187(4)  & 1.84(10) & 0.01(4) & 1.94(10) & 0.10(4) \\
& z=0.7233(4) & & & & \\
\end{tabular}\\
\begin{tabular}{c | c c c c }
\hline
Sym. Op. & $\lbrace m_{yz} | \frac{1}{2} \frac{1}{2} \frac{1}{2} \rbrace$ & $\lbrace m_{xz} | 0 \frac{1}{2} 0 \rbrace$ & $\lbrace m_{xy} | \frac{1}{2} 0 \frac{1}{2} \rbrace$ & $\lbrace 1 | 010 \rbrace$\\
\hline
$\tau_1$ & $\left( \begin{array}{cc} \epsilon & 0 \\ 0 & \epsilon* \end{array} \right)$ &
 $\left( \begin{array}{cc} 0 & \epsilon* \\ \epsilon & 0 \end{array} \right)$ & 
 $\left( \begin{array}{cc} 1 & 0 \\ 0 & 1 \end{array} \right)$ & 
 $\left( \begin{array}{cc} \epsilon^2 & 0 \\ 0 & \epsilon*^2 \end{array} \right)$ \\
\end{tabular}
\caption{Top) Amplitude of the magnetic modulations and relative phases for the four inequivalent Fe sites ($i$=1,4) extracted from the Rietveld refinements in the ICM phase at 30K and CM phase at 1.5K (see text for details of the parametrization). Bottom) Matrix representative of the complete irreducible representation $\tau_1$ for generators of the $Pnma1^\prime$ space group. The symmetry operation are listed in the Seitz notation. Matrices for the time-reversal operator (-Identity) and lattice translations along \textit{a} and \textit{c} (Identity) are not shown. $\epsilon$=e$^{-i\pi\delta}$ and $\epsilon$* denotes the complex conjugate.}
\label{tab:magnetic}
\end{table}
\indent The data obtained at 100K in the paramagnetic regime confirms the crystal structure previously published for the \CaFeAs\ (orthorhombic space group $Pnma$ with a=11.5844 $\AA$, b=3.7368 $\AA$ and c=11.5833 $\AA$). There are four independent Fe sites in the crystallographic unit-cell, all of multiplicity four and in the (x,$\frac{1}{4}$,z) positions and labelled according to Tab. \ref{tab:magnetic}. A small amount of Sn impurity from the flux is also present in the studied sample, and has been treated as a secondary phase in the Rietveld refinements. On cooling below T$_{N1}\approx$85K as shown in Fig. \ref{fig:tdep} , additional Bragg peaks are observed with intensities that decrease rapidly as a function of Q, indicative of long-range magnetic ordering. Based on a grid search \cite{ISI:A1993ME99200007} in the first Brillouin zone, these peaks can be indexed only with an ICM propagation vector k=(0,$\delta$,0) with $\delta$ close to $\frac{3}{8}$. The three most intense peaks are labeled with their reciprocal positions in the thermodiffractograms in Fig. \ref{fig:tdep}. The transition from the paramagnetic phase to this ICM phase is second-order. The value of $\delta$ increases on cooling to reach a maximum close to 0.39 just above T$_{N2}\approx$25K, and changes abruptly below this temperature to lock at the commensurate value $\delta$=$\frac{3}{8}$ within the error of our measurements (Fig. \ref{fig:tdep}). Subsequent measurements with smaller temperature steps close to T$_{N2}$, not shown, indicate that the second transition is first order, with a coexistence of the ICM and CM structures over a temperature regime of 3K. This sequence of magnetic transitions rules out the possible ferromagnetic state investigated by Density Functional Theory\cite{ISI:000265268100019}. Lattice strictions are also observed at the two magnetic transitions, as shown from the temperature dependence of \textit{b} showing a negative thermal expansion below T$_{N1}$ and a small contraction of \textit{c} below T$_{N2}$.\\
\indent Models for the CM and ICM magnetic structures have been derived from Rietveld refinements at 1.5K and 30K, respectively. Due to the large number of independent sites and the limited number of magnetic Bragg peaks, fully unconstrained model even with global optimization procedures do not lead to satisfactory solutions. However, symmetry analysis allows us to reduce greatly the number of possibilities. Since the first transition is second-order, the magnetic order parameter must transform as one of the four one-dimensional irreducible representations of the \textit{Pnma} little group with k=(0,$\delta$,0), labeled $\tau_1$ to $\tau_4$. The systematic absence of the (0,k$\pm \delta$,0) reflections indicate that the moments are aligned along the $\textit{b}$-axis. This is in agreement with the decrease of the magnetic susceptibility along \textit{b} at T$_{N1}$.\\
\begin{figure}[h!] 
\includegraphics[width=210pt]{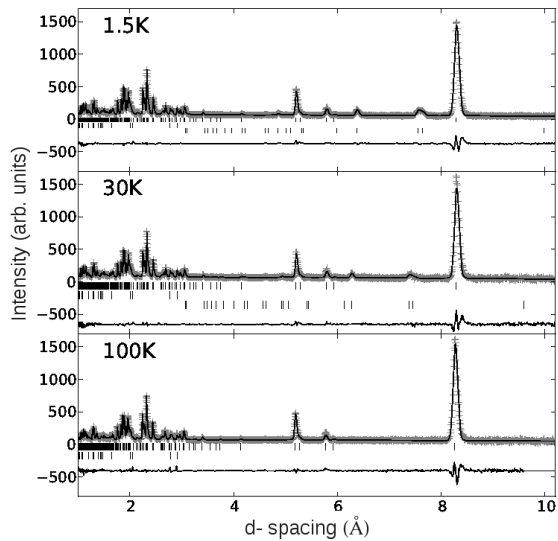} 
\caption{Rietveld refinements of \CaFeAs\ at 1.5 K in the CM phase(top), at 30K in the ICM phase(middle) and 100K in the paramagnetic phase(bottom). On each plot, the gray crosses and solid line represent the experimental data points and calculated diffraction pattern, respectively and the difference is shown at the bottom as a solid line. Rows of markers indicate, from top to bottom, the positions of the nuclear reflections for \CaFeAs\ and Sn (impurity from flux) for all temperatures and the magnetic reflections for \CaFeAs\ in the CM and ICM phase.}
\label{fig:rietveld}
\end{figure}
\indent Within those constraints, we found that only magnetic modes that transform as $\tau_1$ can explain the observed diffraction pattern. In this model, the magnetic structure corresponds to a longitudinal SDW, as postulated in Ref. \cite{ISI:000268617500007}. Each Fe site ($i$=1,4) in position (x,$\frac{1}{4}$,z) listed in Tab. \ref{tab:magnetic} generates three additional Fe positions by symmetry operations of the group with coordinate triplets: (-x,$\frac{3}{4}$,-z),(x+$\frac{1}{2}$,$\frac{1}{4}$,-z+$\frac{1}{2}$),(-x+$\frac{1}{2}$,$\frac{3}{4}$,z+$\frac{1}{2}$). The magnetic moment on each Fe site ($i$=1,4) in the cell R$_L$ is written \textbf{m$_{i}$}=M$_i$ $\hat{j}$ cos(2$\pi($kR$_L$+$\Phi_i$)) where $\hat{j}$ is a unit-vector along b, M$_i$ is the amplitude of the modulation and $\Phi_i$ is the phase, in units of $2\pi$. The $\tau_1$ symmetry constraint requires that the magnetic parameters (M$_i$,$\Phi_i$) transform as (M$_i$,$\Phi_i+\frac{\delta}{2}$), (-M$_i$,$\Phi_i$), (-M$_i$,$\Phi_i+\frac{\delta}{2}$) for these three positions respectively. If no additional physical constraints are imposed, i.e. the amplitude of the modulation is allowed to be different on the inequivalent sites, the magnetic structure is fully described by seven parameters: the four amplitudes M$_i$ and three relative phases, since one of the phase can be fixed to zero (equivalent to an absolute phase for the entire structure). A series of simulated annealing runs with these seven free parameters, using integrated intensities of the magnetic peaks extracted from a Lebail fit, systematically converged to a unique solution. The result was introduced in the Rietveld refinement leading to good agreement with the data and magnetic structure factors of 6.1\%\ and 10.2\%\ at 1.5K and 30K respectively. The value of each refined parameter is shown in Tab. \ref{tab:magnetic} and the Rietveld refinements are displayed in Fig. \ref{fig:rietveld}. We could not exclude the presence of a weaker secondary mode at the ICM-CM transition, since this transition is first-order, but we could not find evidence of additional components below T$_{N2}$ within the statistics and intrinsic limitation of the powder data. In this respect, subsequent work on single crystal diffraction and neutron polarimetry is needed to pin down possible extra modes.\\
\begin{figure}[h!] 
\includegraphics[scale=0.33]{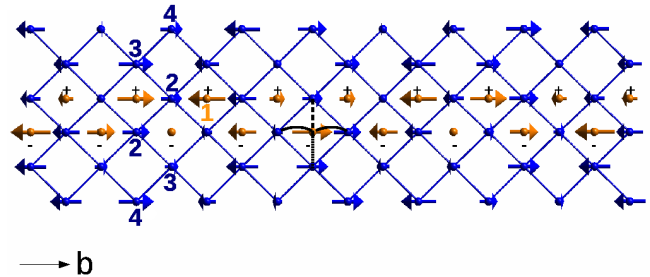} 
\caption{(Color online) Magnetic configuration in the CM phase of \CaFeAs\ at 1.5K along the b-axis. For simplicity, only one Fe$_2$As$_2$ ribbon is shown in blue (dark grey) with capping sites in orange (light grey) and the projected view is perpendicular to the chosen Fe$_2$As$_2$ layer. The + and - signs indicate if the Fe$_1$ ion is located above or below the layer. The Fe sites are labelled 1 to 4 according to the atomic positions given in Tab. \ref{tab:magnetic}. The As atoms sublattice is not shown for clarity. Straight solid lines mark the connectivity in the Fe$_2$As$_2$ layer. Three inequivalent magnetic exchange paths between five-coordinated Fe$_1$ and atoms in the Fe$_2$As$_2$ layer are shown respectively as short-dashed, long-dashed and curled lines.}
\label{fig:structure}
\end{figure}   
\indent All magnetic sites have an ordered magnetic component in the ICM phase with moments smaller than 2$\mu_B$. The most remarkable change at the ICM-CM transition, in addition to the lock-in of the wave-vector, is an increase of the amplitude on the Fe$_1$ site, i.e. the five-coordinated site connecting adjacent Fe$_2$As$_2$ ribbons. It is however impossible to determine whether the small change of relative phases detected at the CM transition is real, for which precise refinements of the magnetic parameters on single crystal data will be necessary. At 1.5K, the modulation amplitude is the largest on Fe$_1$, and the lowest on Fe$_2$. This is also what was found by Density Functional Theory (DFT) calculations \cite{ISI:000265268100019}, although they were performed assuming a ferromagnetic state. Indeed, the ordered moment on the five-coordinated Fe was calculated to be 2.06 $\mu_B$ and the smallest moment was 1$\mu_B$ on Fe$_2$ (labelled Fe$_3$ in \cite{ISI:000265268100019}). M\"{o}ssbauer spectroscopy also suggests that out of the four iron sites, only the five-coordinated site has the formal oxidation state Fe$^+$, which is consistent with the higher moment. The magnetic arrangement is shown in Fig. \ref{fig:structure} for a single Fe$_2$As$_2$ ribbon capped by first-neighbour pyramidal Fe$_1$ sites.\\
\indent The origin of the modulation can be understood within a localized picture by considering and known magnetic configuration of the parent 122 compound and its analogs \cite{ISI:000261891200064,ISI:000256632000038}. In such systems, ferromagnetic rows in one direction of the square lattice are coupled antiferromagnetically along the other direction, leading to anti-parallel spins along the square diagonals (AFM next-nearest exchange). In \CaFeAs\, additional exchange couplings are introduced between the layers and Fe$_1$ sites. There are three different exchange paths marked by different symbols in Fig. \ref{fig:structure}. Importantly, the exchange integrals between Fe$_1$ and Fe$_2$ along the \textit{b}-axis are equivalent by symmetry (left and right) due to the mirror plane perpendicular to \textit{b}. This implies that whatever the sign of this exchange interaction, it will compete with the in-plane next-nearest neighbor exchange. This is topologically equivalent to a chain with competing first- and second-neighbor interactions and can explain the long-wavelength modulation along the \textit{b}-axis. Along the other diagonal, competition is also present but the two exchange integrals are not equivalent by symmetry. Although there is a modulation along this diagonal inside a single Fe$_2$As$_2$ ribbon (Fig. \ref{fig:structure}), the presence of four ribbons per unit-cell interconnected through Fe$_1$  sites on a non-frustrated cross-pattern lattice, does not give rise overall to a modulation along \textit{a} or \textit{c}.\\ 
\indent Finally, with the knowledge of the magnetic phase transitions, we suggest possible mechanisms for the observed  anomalies in the heat capacity and resistivity by applying Landau theory. Taking into account the symmetry of the two-component magnetic order-parameter $\eta$ = $\mid\eta\mid$ e${^{i\varphi} }$ and $\eta$ ${^*}$= $\mid \eta \mid$ e${^{-i\varphi}}$ transforming as $\tau_1$(\ref{tab:magnetic}) one gets the Landau free-energy density
\begin{equation}
\Phi (T,\mid\eta\mid,\varphi)=\Phi_0(T)+\frac{\alpha}{2}\eta^2+\frac{\beta}{4}\eta^4+\frac{\gamma_1}{16}\eta^{16}cos(16\varphi) 
\label{eq:FreeEnergy}
\end{equation}
where $\gamma_1$=0 at the transition to the ICM phase, which displays the \textit{mmm1'} magnetic point-group symmetry, and $\gamma_1$ $\neq$ 0 at the transition to the CM phase. Minimization of $\Phi$ yields three possible stable structures for the CM phase, having the respective symmetries \textit{P112$_1$/a}($\varphi$=n$\pi$/8), \textit{P2$_1$ma}($\varphi$=(2n+1)$\pi$/16) and \textit{P11a}($\varphi\neq$n$\pi$/16) , with an eightfold magnetic unit-cell. From the refinement of the crystal structure, there is no indication of peak broadening at the CM-ICM phase, which suggests that the CM phase retains an orthorhombic symmetry (\textit{P2$_1$ma}). Anyhow, for all the aforementioned symmetries, the sixteenth degree term of the lock-in invariant, corresponding to k=(0,$\frac{3}{8}$,0), is key to explain the coupling to the physical properties and describe the unusual temperature dependences of the specific heat C, which shows no noticeable anomaly at T$_{N2}$ and the resistivity $\rho$, which exhibits a sharp drop below T$_{N2}$. Since the $\gamma_1$ invariant induces in the CM phase a higher order perturbation $\Delta\eta^2$ $\approx$ $\gamma_1$ (T$_{N2}$-T)$^{\frac{1}{7}}$ with respect to its equilibrium value $\eta^2$=$\frac{\alpha_0}{\beta}$ (T$_{N1}$-T) in the ICM phase, the specific heat C = -T$\frac{\partial \Phi^2}{\partial T^2}$ shows, on cooling below T$_{N2}$, a slight drop in the slope $\Delta C \propto T(T_{N2}-T)^{\frac{-6}{7}}$ with respect to its linear temperature dependence $\Delta C=\frac{\alpha_0^2}{2\beta}$T above T$_{N2}$, as reported experimentally. The resistivity contribution to the free-energy $\Phi^\rho$=$\mu\eta^2\rho$+$\frac{\nu}{2}\rho^2 $ gives in the ICM phase  $\rho^e$(T)=-$\frac{\mu}{\nu}\eta^2$= -$\frac{\mu\alpha_0}{\nu\beta}$(T$_{N1}$-T) corresponding to a linear decrease of $\rho^e$(T). In the CM phase, the coupling of $\rho$ to the additional sixteenth-degree lock-in term yields $\rho^e$(T)= -$\frac{\alpha_0}{\nu\beta}$ [$\mu$(T$_{N2}$-T)+$\mu$'(T$_{N2}$-T)$^8$]. This  corresponds to a strong drop of $\rho^e$(T) below T$_{N2}$ assuming a large positive value of the coupling coefficient $\mu$'.\\
\indent In summary, we have presented for the first time evidence for two magnetic transitions in \CaFeAs\ corresponding to longitudinal spin-density-waves, incommensurate below T$_{N1}$=85K and locked-in at k=(0,$\frac{3}{8}$,0) below T$_{N2}$=25K, as determined by neutron powder diffraction experiments. Models for the magnetic structures in both phases have been derived from symmetry considerations and Rietveld refinements of the neutron data. The lock-in transition can be explained phenomenologically by the presence of a sixteenth degree invariant in the thermodynamic potential that reproduces the anomaly observed in the temperature dependence of the electrical resistivity. An in-depth analysis of the microscopic origin of these two magnetic states would require first principle calculations, which could also indicate the conditions under which superconductivity could be introduced in this compound.\\ 
\indent Research at Argonne National Laboratory is supported by the U.S. Department of Energy, Office of Basic Energy Sciences, Division of Materials Sciences and Engineering under Award No. DE-AC02-06CH11357.\\


\begin{thebibliography}{9}
\expandafter\ifx\csname natexlab\endcsname\relax\def\natexlab#1{#1}\fi
\expandafter\ifx\csname bibnamefont\endcsname\relax
  \def\bibnamefont#1{#1}\fi
\expandafter\ifx\csname bibfnamefont\endcsname\relax
  \def\bibfnamefont#1{#1}\fi
\expandafter\ifx\csname citenamefont\endcsname\relax
  \def\citenamefont#1{#1}\fi
\expandafter\ifx\csname url\endcsname\relax
  \def\url#1{\texttt{#1}}\fi
\expandafter\ifx\csname urlprefix\endcsname\relax\def\urlprefix{URL }\fi
\providecommand{\bibinfo}[2]{#2}
\providecommand{\eprint}[2][]{\url{#2}}

\bibitem[{\citenamefont{Kamihara et~al.}({2008})\citenamefont{Kamihara,
  Watanabe, Hirano, and Hosono}}]{ISI:000253951900032}
\bibinfo{author}{\bibfnamefont{Y.}~\bibnamefont{Kamihara}},
  \bibinfo{author}{\bibfnamefont{T.}~\bibnamefont{Watanabe}},
  \bibinfo{author}{\bibfnamefont{M.}~\bibnamefont{Hirano}}, \bibnamefont{and}
  \bibinfo{author}{\bibfnamefont{H.}~\bibnamefont{Hosono}},
  \bibinfo{journal}{{J. Am. Chem. Soc.}} \textbf{\bibinfo{volume}{{130}}},
  \bibinfo{pages}{{3296}} (\bibinfo{year}{{2008}}).

\bibitem[{\citenamefont{Huang et~al.}({2008})\citenamefont{Huang, Qiu, Bao,
  Green, Lynn, Gasparovic, Wu, Wu, and Chen}}]{ISI:000261891200064}
\bibinfo{author}{\bibfnamefont{Q.}~\bibnamefont{Huang}},
  \bibinfo{author}{\bibfnamefont{Y.}~\bibnamefont{Qiu}},
  \bibinfo{author}{\bibfnamefont{W.}~\bibnamefont{Bao}},
  \bibinfo{author}{\bibfnamefont{M.~A.} \bibnamefont{Green}},
  \bibinfo{author}{\bibfnamefont{J.~W.} \bibnamefont{Lynn}},
  \bibinfo{author}{\bibfnamefont{Y.~C.} \bibnamefont{Gasparovic}},
  \bibinfo{author}{\bibfnamefont{T.}~\bibnamefont{Wu}},
  \bibinfo{author}{\bibfnamefont{G.}~\bibnamefont{Wu}}, \bibnamefont{and}
  \bibinfo{author}{\bibfnamefont{X.~H.} \bibnamefont{Chen}},
  \bibinfo{journal}{{Phys. Rev. Lett.}} \textbf{\bibinfo{volume}{{101}}}
  (\bibinfo{year}{{2008}}).

\bibitem[{\citenamefont{Zhi-An et~al.}({2008})\citenamefont{Zhi-An, Wei, Jie,
  Wei, Xiao-Li, Zheng-Cai, Guang-Can, Xiao-Li, Li-Ling, Fang
  et~al.}}]{ISI:000256252600080}
\bibinfo{author}{\bibfnamefont{R.}~\bibnamefont{Zhi-An}},
  \bibinfo{author}{\bibfnamefont{L.}~\bibnamefont{Wei}},
  \bibinfo{author}{\bibfnamefont{Y.}~\bibnamefont{Jie}},
  \bibinfo{author}{\bibfnamefont{Y.}~\bibnamefont{Wei}},
  \bibinfo{author}{\bibfnamefont{S.}~\bibnamefont{Xiao-Li}},
  \bibinfo{author}{\bibfnamefont{L.}~\bibnamefont{Zheng-Cai}},
  \bibinfo{author}{\bibfnamefont{C.}~\bibnamefont{Guang-Can}},
  \bibinfo{author}{\bibfnamefont{D.}~\bibnamefont{Xiao-Li}},
  \bibinfo{author}{\bibfnamefont{S.}~\bibnamefont{Li-Ling}},
  \bibinfo{author}{\bibfnamefont{Z.}~\bibnamefont{Fang}}, \bibnamefont{et~al.},
  \bibinfo{journal}{{Ch. Phys. Lett.}} \textbf{\bibinfo{volume}{{25}}},
  \bibinfo{pages}{{2215}} (\bibinfo{year}{{2008}}).

\bibitem[{\citenamefont{Rotter et~al.}({2008})\citenamefont{Rotter, Tegel, and
  Johrendt}}]{ISI:000258975100055}
\bibinfo{author}{\bibfnamefont{M.}~\bibnamefont{Rotter}},
  \bibinfo{author}{\bibfnamefont{M.}~\bibnamefont{Tegel}}, \bibnamefont{and}
  \bibinfo{author}{\bibfnamefont{D.}~\bibnamefont{Johrendt}},
  \bibinfo{journal}{{Phys. Rev. Lett.}} \textbf{\bibinfo{volume}{{101}}}
  (\bibinfo{year}{{2008}}).

\bibitem[{\citenamefont{Todorov et~al.}({2009})\citenamefont{Todorov, Chung,
  Malliakas, Li, Bakas, Douvalis, Trimarchi, Gray, Mitchell, Freeman
  et~al.}}]{ISI:000265268100019}
\bibinfo{author}{\bibfnamefont{I.}~\bibnamefont{Todorov}},
  \bibinfo{author}{\bibfnamefont{D.~Y.} \bibnamefont{Chung}},
  \bibinfo{author}{\bibfnamefont{C.~D.} \bibnamefont{Malliakas}},
  \bibinfo{author}{\bibfnamefont{Q.}~\bibnamefont{Li}},
  \bibinfo{author}{\bibfnamefont{T.}~\bibnamefont{Bakas}},
  \bibinfo{author}{\bibfnamefont{A.}~\bibnamefont{Douvalis}},
  \bibinfo{author}{\bibfnamefont{G.}~\bibnamefont{Trimarchi}},
  \bibinfo{author}{\bibfnamefont{K.}~\bibnamefont{Gray}},
  \bibinfo{author}{\bibfnamefont{J.~F.} \bibnamefont{Mitchell}},
  \bibinfo{author}{\bibfnamefont{A.~J.} \bibnamefont{Freeman}},
  \bibnamefont{et~al.}, \bibinfo{journal}{{J. Am. Chem. Soc.}}
  \textbf{\bibinfo{volume}{{131}}}, \bibinfo{pages}{{5405}}
  (\bibinfo{year}{{2009}}).

\bibitem[{\citenamefont{Zhao et~al.}({2009})\citenamefont{Zhao, Yi, Fettinger,
  Kauzlarich, and Morosan}}]{ISI:000268617500007}
\bibinfo{author}{\bibfnamefont{L.~L.} \bibnamefont{Zhao}},
  \bibinfo{author}{\bibfnamefont{T.}~\bibnamefont{Yi}},
  \bibinfo{author}{\bibfnamefont{J.~C.} \bibnamefont{Fettinger}},
  \bibinfo{author}{\bibfnamefont{S.~M.} \bibnamefont{Kauzlarich}},
  \bibnamefont{and} \bibinfo{author}{\bibfnamefont{E.}~\bibnamefont{Morosan}},
  \bibinfo{journal}{{Phys. Rev. B}} \textbf{\bibinfo{volume}{{80}}}
  (\bibinfo{year}{{2009}}).

\bibitem[{\citenamefont{Rodriguez~Carvajal}({1993})}]{ISI:A1993ME99200007}
\bibinfo{author}{\bibfnamefont{J.}~\bibnamefont{Rodriguez~Carvajal}},
  \bibinfo{journal}{{Physica B}} \textbf{\bibinfo{volume}{{192}}},
  \bibinfo{pages}{{55}} (\bibinfo{year}{{1993}}).

\bibitem[{\citenamefont{Kovalev}(1993)}]{rep:kovalev}
\bibinfo{author}{\bibfnamefont{O.~V.} \bibnamefont{Kovalev}},
  \emph{\bibinfo{title}{Representations of the Crystallographic Space Groups}}
  (\bibinfo{publisher}{Edition 2, Gordon and Breach Science Publishers},
  \bibinfo{address}{Switzerland}, \bibinfo{year}{1993}).

\bibitem[{\citenamefont{de~la Cruz et~al.}({2008})\citenamefont{de~la Cruz,
  Huang, Lynn, Li, Ratcliff, Zarestky, Mook, Chen, Luo, Wang
  et~al.}}]{ISI:000256632000038}
\bibinfo{author}{\bibfnamefont{C.}~\bibnamefont{de~la Cruz}},
  \bibinfo{author}{\bibfnamefont{Q.}~\bibnamefont{Huang}},
  \bibinfo{author}{\bibfnamefont{J.~W.} \bibnamefont{Lynn}},
  \bibinfo{author}{\bibfnamefont{J.}~\bibnamefont{Li}},
  \bibinfo{author}{\bibfnamefont{W.}~\bibnamefont{Ratcliff},
  \bibfnamefont{II}}, \bibinfo{author}{\bibfnamefont{J.~L.}
  \bibnamefont{Zarestky}}, \bibinfo{author}{\bibfnamefont{H.~A.}
  \bibnamefont{Mook}}, \bibinfo{author}{\bibfnamefont{G.~F.}
  \bibnamefont{Chen}}, \bibinfo{author}{\bibfnamefont{J.~L.}
  \bibnamefont{Luo}}, \bibinfo{author}{\bibfnamefont{N.~L.}
  \bibnamefont{Wang}}, \bibnamefont{et~al.}, \bibinfo{journal}{{Nature}}
  \textbf{\bibinfo{volume}{{453}}}, \bibinfo{pages}{{899}}
  (\bibinfo{year}{{2008}}).
\end{thebibliography}

\end{document}